\documentclass{emulateapj}
\usepackage{times}
\usepackage{graphicx}

\slugcomment{Submitted to ApJ }

\shorttitle{IBS Heating of Black Widows}
\shortauthors{Romani \& Sanchez}

\begin{document}

\title{Intra-Binary Shock Heating of Black Widow Companions}

\author{Roger W. Romani \& Nicolas Sanchez\altaffilmark{1}}
\altaffiltext{1}{Department of Physics, Stanford University, Stanford, CA 94305-4060,
 USA; rwr@astro.stanford.edu}

\begin{abstract}

	The low mass companions of evaporating binary pulsars (black widows and their ilk) 
are strongly heated on the side facing the pulsar. However in high-quality photometric and 
spectroscopic data the heating pattern does not match that expected for direct pulsar 
illumination. Here we explore heating mediated by an intra-binary shock (IBS). We develop a simple
analytic model and implement it in the popular `ICARUS' light curve code. The model is parameterized
by the wind momentum ratio $\beta$ and velocity $v_{\rm Rel}v_{\rm orb}$ and assumes that the 
reprocessed pulsar
wind emits prompt particles or radiation to heat the companion surface. We illustrate an 
interesting range of light curve asymmetries controlled by these parameters. The code also computes
the IBS synchrotron emission pattern, and thus can model black widow X-ray light curves. As a 
test we apply the results to the high quality asymmetric optical light curves of PSR J2215+5135; 
the resulting fit gives a substantial improvement upon direct heating models and produces an
X-ray light curve consistent with that seen. The IBS model parameters imply that, at the present loss
rate, the companion evaporation has a characteristic timescale $\tau_{\rm evap} \approx 150$My.
Still, the model is not fully satisfactory, indicating additional unmodeled physical effects.
\end{abstract}

\keywords{gamma rays: stars --- pulsars: general}

\section{Introduction}

	Since discovery of the original `black widow' pulsar PSR J1959+2048, it has been 
realized that optical study of the pulsar companion provides an important path to understanding
the dynamics of these exotic systems, including the pulsar heating mechanism, the companion
wind and the component masses \citep{de88,arc92,cvr95}. Such studies became even more interesting when 
\citet{vKBK11}
found evidence that this pulsar might be especially massive, so that precision
measurements of black widow component masses could have important implication 
for our understanding of the binary evolution and of the dense matter equation of state.

	The quest seems straight-forward. Radio or gamma-ray pulsar timing provides
a precise orbital ephemeris and a companion mass function, via the orbital period $P_B$
and projected semi-major axis of the pulsar orbit $x=a_p {\rm sin}i$:
$$
f(M_P,q,i)={{4\pi^2x^3} \over {G P_B^2}}=M_P {{({\rm sin}i)^3} \over {q(q+1)^2}}
$$
with $M_P$ and $M_C$ the pulsar and companion masses and $q=M_P/M_C$ the mass ratio.
In clean double-degenerate systems, relativistic effects in precision pulsar timing 
allow solution for $q$ and $i$. For the companion-evaporating pulsars (black widows, with
$M_C \approx 0.01-0.03M_\odot$, redbacks with $M_C \approx 0.1-0.3M_\odot$, and their ilk)
the dissipation and propagation effects of the companion and wind preclude such precision timing.
However, optical studies of the companion can, in principle, measure the spectroscopic 
radial velocity amplitude $K_C$ (giving $q=K_C P_B/2\pi x$) and, by measuring the 
optical modulation due to varying view of the heated side, the orbital inclination $i$.

The challenge is that the radial velocity observed is weighted toward the center of light on the heated side
so that $K_C=K_{obs} K_{corr}$ is larger that the observed radial velocity amplitude by
$K_{cor} \approx 1.03-1.08\times$ \citep{rfc15}, depending on $i$ and the heating pattern. One 
commonly assumes that the pulsar spindown power heats the facing side of the companion 
directly, raising the characteristic temperature from the unheated (`Night' side) $T_N$ to
$$
T_D^4=\eta {\dot E}/4\pi a^2 \sigma +T_N^4
$$
with $a=x(1+q)/{\rm sin}i$ the orbital separation, ${\dot E}=I\Omega{\dot \Omega}$ the
pulsar spindown power for moment of inertia $I$ and $\eta$ a heating efficiency.
This model has been implemented in several light curve modeling codes
eg. the ELC code \citep{oh00} and its descendant ICARUS \citep{bet13}.
Direct isotropic `photon' heating is assumed, which is indeed a good approximation for many X-ray binaries.
Fitting black widow light curves and spectra with such codes has led to surprisingly
large estimates of $M_N$: $2.4\pm0.12 M_\odot$ for PSR J1959+2048 \citep{vKBK11} and
$\approx 2.7 M_\odot$ for PSR J1311$-$3430 \citep{rfc15}.

	However, with the discovery of a large population of BW and RB in the direction of 
{\it Fermi} sources, several nearby, bright systems have been found, enabling
high precision optical light curves and spectroscopy. Direct heating models, 
which adequately described some early low precision observations, often do not provide a 
statistically acceptable description of the high precision data.
In particular, many light curves are substantially asymmetric \citep{stap01,sh14}, the color and
spectral variations across the face do not match direct heating patterns \citep{rfc15} and
the inferred heating power in several cases show large $\eta \ge 1$. This implies
that the pulsar power does not heat the companion via direct illumination but that pulsar 
particles or high energy radiation are deflected before reaching the companion.
In one natural scenario the pulsar and companion winds set up an intra-binary shock (IBS);
the heating power arises in this structure. In fact the X-ray light curves of many BW and RB
show modulation indicating such IBS \citep{robet14}. A second plausible picture invokes 
a companion magnetic field intercepting the pulsar wind and channeling spindown power 
to the surface. Since companion magnetic structures are at present poorly constrained,
we focus here on a 
pseudo-analytic model for IBS-mediated illumination, allowing robust fits for the principal
physical wind parameters in data fitting codes.
We have developed an ICARUS module employing this model. The results can mimic
a range of observed BW/RB behavior and result in dramatically improved light curve fits. 
However, we show that some aspects are not adequately modeled and close by briefly
describing additional physical ingredients, such as companion fields, likely needed in the fits.

\begin{figure}[t!!]
\vskip 5.5truecm
\includegraphics{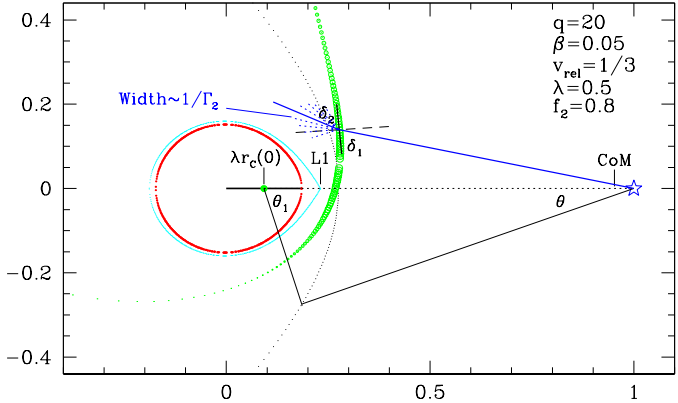}
\begin{center}
\caption{\label{spectra} 
Intra-Binary Shock geometry in the orbital plane, here for a mass ratio $q=20$ (RB-like), 
a wind momentum ratio $\beta=0.05$
and a relative velocity $v_{\rm Rel}=1/3$. The wind origin is placed halfway toward the companion
`nose' facing the $L_1$ point. The wind angles $\theta_1,\, \theta_2$ are shown for 
the $v_{\rm Rel}=\infty$ case, without
sweep-back. The local intensity of the swept shock surface is indicated by point size and
the shock inclination angles and post-shock beaming are shown for one ray.
}
\end{center}
\vskip -0.5truecm
\end{figure}

\section{Intra-Binary Shock Model}

Although existing codes such as ICARUS assume isotropic pulsar irradiation, the relativistic
wind is likely equatorially concentrated as $\propto {\rm sin}^2\theta$, with $\theta$
the polar angle
$$
f_P(r,\theta) = {\dot E(\theta)}/(4\pi r^2c) = 3I\Omega{\dot \Omega}{\rm sin}^2\theta/(4\pi r^2 c).
$$
In fact \citet{tsl13} find that, for large pulsar inclination angle $\alpha$, numerical 
simulations show a wind power distributed as approximately ${\rm sin}^4\theta$. We will
assume a quadratic form here, and compare with the $\theta^0$ and $\theta^4$ cases.
This wind shocks against a baryonic companion wind of speed $v_W$ and mass loss rate ${\dot M}_W$,
which gives rise to a momentum flux 
$$
f_C(r)={\dot M}_W v_W/(4\pi r^2).
$$
We will assume that this wind is isotropic. Thus the wind shock geometry is controlled by two
principal parameters, the wind momentum flux ratio
$$
\beta =  {\dot M}_W v_W c/{\dot E}
$$
and the ratio between the massive wind speed and the orbital velocity $v_{\rm Rel} = v_W/v_{orb}$.
The companion wind is driven (in a poorly understood way) by pulsar irradiation. Further, in
the Roche geometry the escape potential is lowest at the L1 point. These effects suggest a 
companion wind centered inward of its center of mass. We parametrized this shift 
with a secondary parameter $\lambda$, with $\lambda=0$ for a wind centered on the star and
$\lambda=1$ centered on the star surface at the sub-pulsar point closest to L1 ($r_C(0)$
from the companion center). This parameter does not have a strong effect unless the 
companion wind is quite weak (small $\beta$).

	We compute the contact discontinuity surface (implicitly assuming a thin
shell IBS, with rapid cooling). The wind origins are at the pulsar and a distance 
$d=a-\lambda r_C(0)$ toward the companion along the binary axis.  This locates the
IBS as a surface of revolution about the binary axis with 
$$
r(\theta) = d ~ {\rm sin}\theta_1/{\rm sin}(\theta + \theta_1)
$$
\citep{crw96}, where $\theta$ is the angle between the line of centers and the ray from
the pulsar and
$$
\theta_1=\left [ {15\over 2} \left ( \left [
1+{4\over 5}\beta (1-\theta {\rm cot}\theta)\right ]^{1/2} -1 \right ) \right ]^{1/2}
$$
is the equivalent angle from the companion wind center (Figure 1). This describes the intersection
of two stationary winds. In our case the motion of the companion causes the shock symmetry
axis to trace an Archimedean spiral \citep{pp08}, lagged in true anomaly behind the 
center of mass position by an orbital phase angle $\delta \phi_B(r) \approx r/(2\pi d v_{\rm Rel})$. 
The resulting geometry compares well with shock structures seen in numerical simulations
\citep[e.g.][]{bbp15}
and the two parameter family captures the range of wind ratios and orbital distortion,
while remaining quickly calculable.

	The pulsar wind has a transverse embedded field, whose magnetization parameter
$\sigma= B^2/(4\pi \gamma \rho c^2)$ is poorly known. Without reconnection, the transverse
toroidal field should have a value $B(r) \approx 4 B_P (2 \pi/cP)^2 r_{NS}^3/r$ in
the post shock region.  For a typical MSP surface dipole $B=10^8B_8$\,G spin period
$P=3P_3$\,ms and orbital separation $a\sim 10^{11}$cm, this 
is $B_{IBS}\approx 20 B_8 a_{11}^{-1} P_3^{-2}$\,G.
Away from the sub-pulsar point the shock is oblique with
an angle to the shock decreasing from $\delta_1$ to a post-shock
$\delta_2 = {\rm tan}^{-1} \chi(\sigma) \delta_1$ (Figure 1) with 
$$
\chi(\sigma)= [1+2\sigma+(16\sigma^2+16\sigma +1)^{1/2}]/(6+6\sigma)
$$
and post-shock bulk $\Gamma_2 = (1-\chi^2)^{-1/2}/{\rm sin}\delta_1$
\citep{kl11}. These relations hold for ultra-relativistic transverse field flow, even
when the field is not in the shock plane (Y. Yuan, private communication). Given our 
poor knowledge of the wind properties we assume $\sigma=1$ in the following. We have 
explored the magnetization dependence; in principle a well understood shock geometry
allows a probe of this important parameter, but we will not discuss that dependence here.

	In the spirit of our thin shock approximation, we assume prompt radiation
from the shocked energetic pulsar wind. We take this to imply radiation in a Gaussian beam
of width $\sigma_r=1/\Gamma_2$ centered around the immediate post-shock flow vector
$\pi/2 - \delta_2$ to the local shock normal. If we assume that the full power of the 
pulsar wind impacting on the IBS $f_p(r,\theta)$ is promptly re-radiated from 
$e^\pm$ at large pitch angle, it emits
$\approx f_p(r,\theta) {\rm cos}(\pi/2-\delta_1)/\Gamma_2$ of the wind energy per unit area.
With the $B_{IBS}$ above, the bright IBS X-rays come from particles with
$\gamma \approx 5 \times 10^4 (E_{keV}/B_{IBS})^{1/2}$ and a cooling time 
$\tau \approx 30 E_{keV}^{-1/2} B_{IBS}^{-3/2}$\,s. With $B_8 \sim 2-3$ this is comparable
to the flow time near the stagnation point at the IBS apex where 
$t_{flow} > a/(c/3) \sim 10$\,s.  
Note that the shock is strong at the shock apex, so that $\Gamma_2$ is small and the 
radiation is widely beamed. In contrast, as the shock becomes tangential down stream
(or as swept back by orbital motion) $\Gamma_2$ grows. Thus both the solid angle
and shock weakening decrease the prompt-emission surface brightness, although this
decreased emission is increasingly beamed closer to the shock limb.

\begin{figure*}[t!!]
\vskip 11.0truecm
\includegraphics{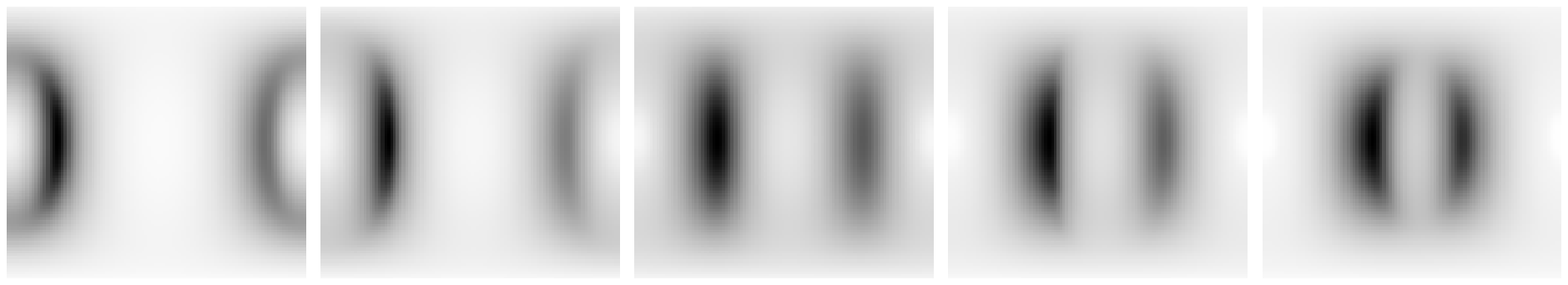}
\includegraphics{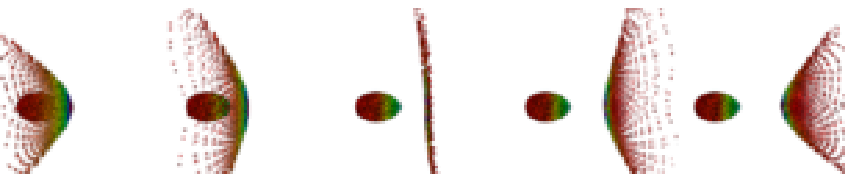}
\includegraphics{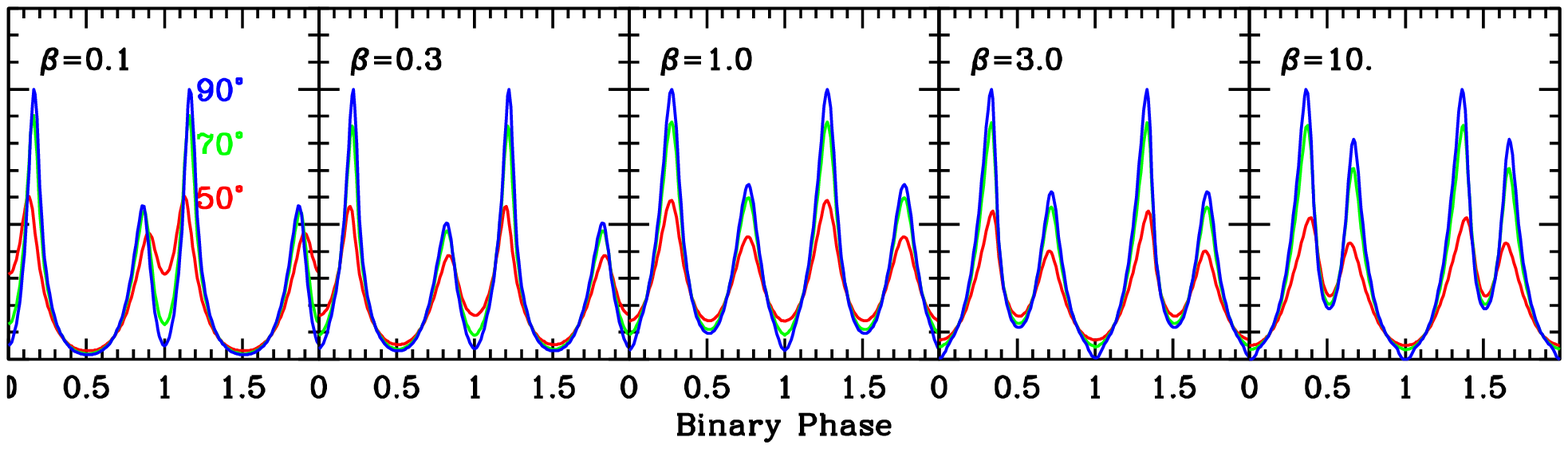}
\begin{center}
\caption{\label{X-rayCs} 
IBS dependence on the wind momentum flux ratio, here for $v_{Rel}=3$. 
Left to right: small $\beta$ (pulsar dominated)
to large $\beta$ (companion dominated). Top: images of the heated companion and the IBS
shock contact discontinuity, viewed from the orbital plane at $\phi_B=\pi/2$. 
Middle: IBS Synchrotron emission pattern on the sky, with each panels covering $\phi_B=0$
(pulsar inferior conjunction) to $\phi_B=2\pi$ and viewing angle $i=0$ to $\pi$. Bottom: IBS
synchrotron light curves for the sample inclinations $i=90^\circ,\,70^\circ,\, 50^\circ$; two orbital periods are 
shown for clarity.
}
\end{center}
\vskip -0.5truecm
\end{figure*}

\begin{figure}[h!!]
\vskip 3.9truecm
\includegraphics{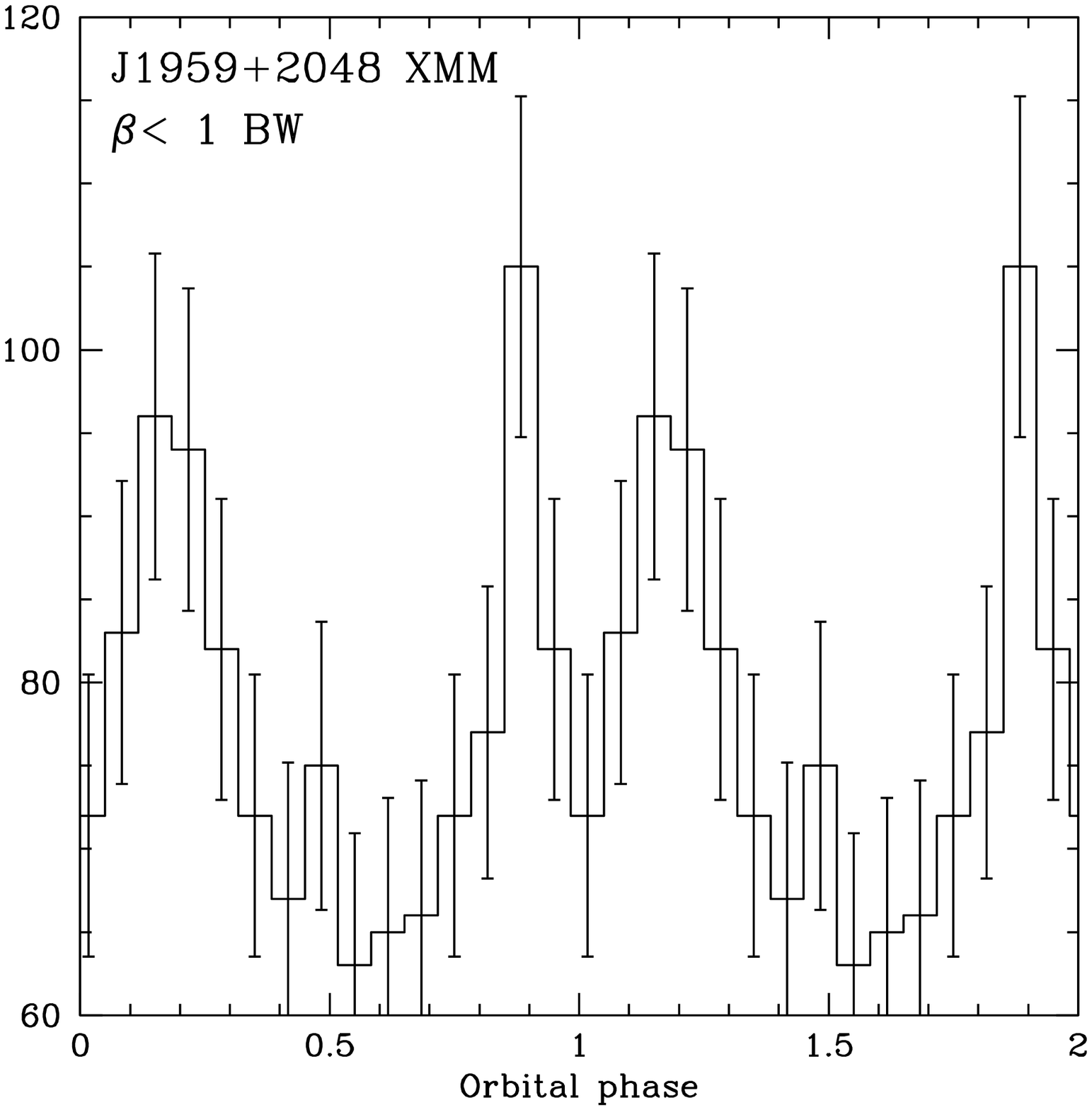}
\includegraphics{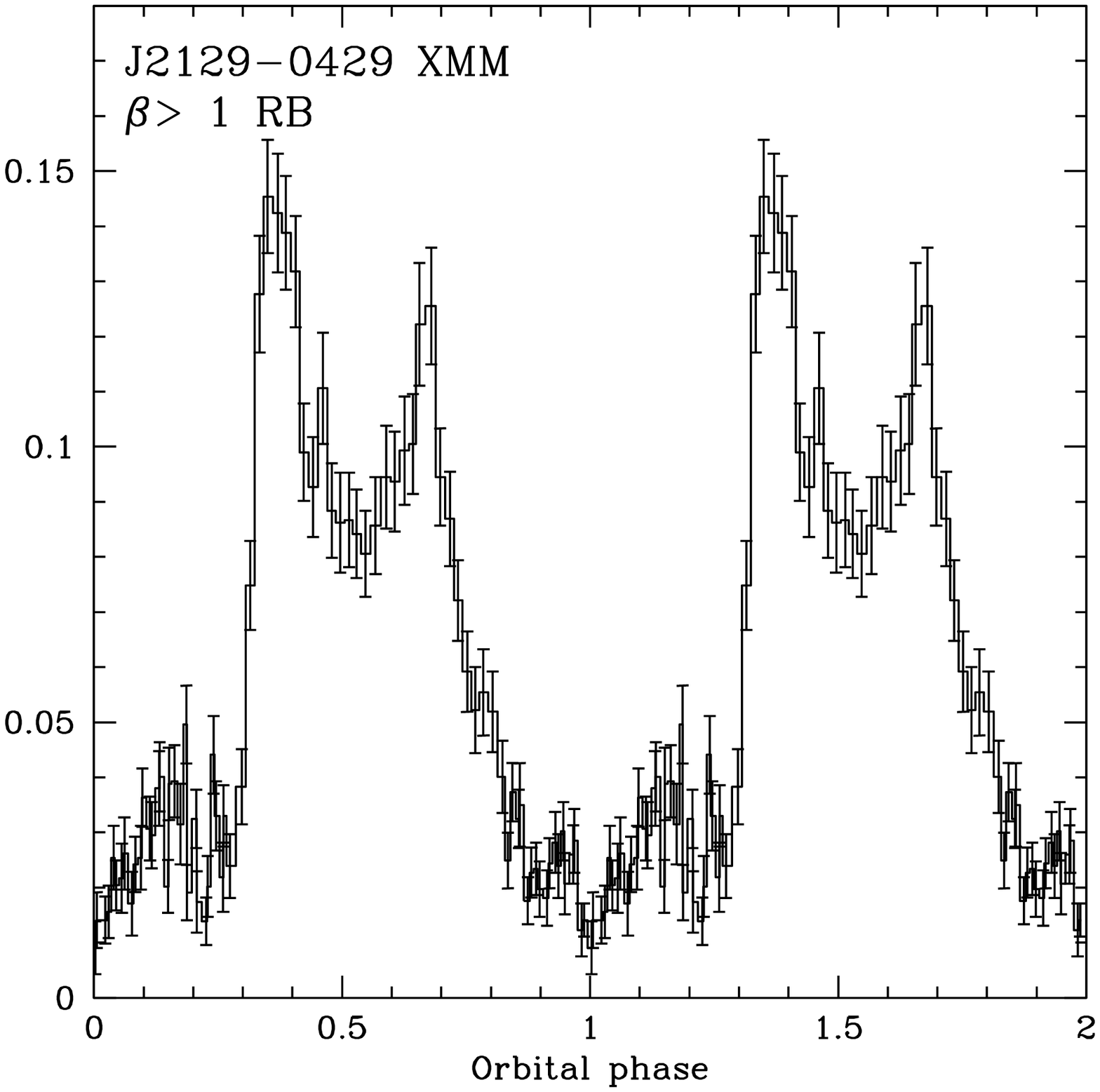}
\begin{center}
\caption{\label{ShsurfLCs} 
Example XMM X-ray light curves, with two periods and phasing as in the lower row of
Figure 2. J1959+2048 is a powerful black widow with weak companion wind (low $\beta$).
PSR J2129$-$0429 is a low power redback undergoing qRLOF (likely high $\beta$).
These can be compared with the first and last panels of the bottom row of figure 2.
}
\end{center}
\vskip -0.5truecm
\end{figure}

	Some post-shock radiation will not be prompt. Indeed most may be slow
e.g. if strong reconnection drives $\sigma \ll 1$, reducing $B_{IBS}$, or if the immediate post-shock 
pitch angle is small. In this case the radiation persists as the shocked wind flows
a distance $\sim a$. Then we may assume that the emission is directed approximately
tangent to the contact discontinuity, and that the shocked pulsar wind accelerates
as it flows away from the shock apex. Examining numerical simulations of \citet{boget12} we
can approximate the resulting bulk $\Gamma_\parallel$ as
$$
\Gamma_\parallel \approx 1.2(1+dr/r_0)
$$
where $r_0$ is the standoff distance (along the line of centers) of the weaker wind and
for a given position on the IBS $dr$ is the increase in radial distance from the stronger
wind center to that along the line of centers (i.e. $dr=0$ at the nose).
The numerical simulations do not give a clear prescription for the emissivity;
the particle density drops rapidly behind the apex, but the magnetic field appears to initially
grow in the post shock flow before downstream dilution. For simplicity and to compare with
the prompt emission scheme, we assign surface brightness as above, scaled with the diminishing pulsar
wind flux per unit IBS area; this mimics the downstream fading expected from the simulations.

	These expressions for the shock emissivity and its re-radiation heating of the
companion surface have been implemented in the ICARUS code. This includes options to
compute the optical companion light curve and the synchrotron IBS-dominated light curves
for any desired inclination. The code also minimizes residuals with respect to optical data
to determine model parameters and errors. While assembling these routines, we noted that the
standard ICARUS distribution was missing a ${\rm cos} \chi$ (with $\chi$ the angle to the 
local surface normal) in the companion heating computation. There was also an error in the
treatment of limb darkening. These have now been amended, but
parameters fit with ICARUS before 2016 will likely need updating. The IBS modules and 
instructions for their inclusion will be posted on Github.

\begin{figure*}[t!!]
\vskip 8.5truecm
\includegraphics{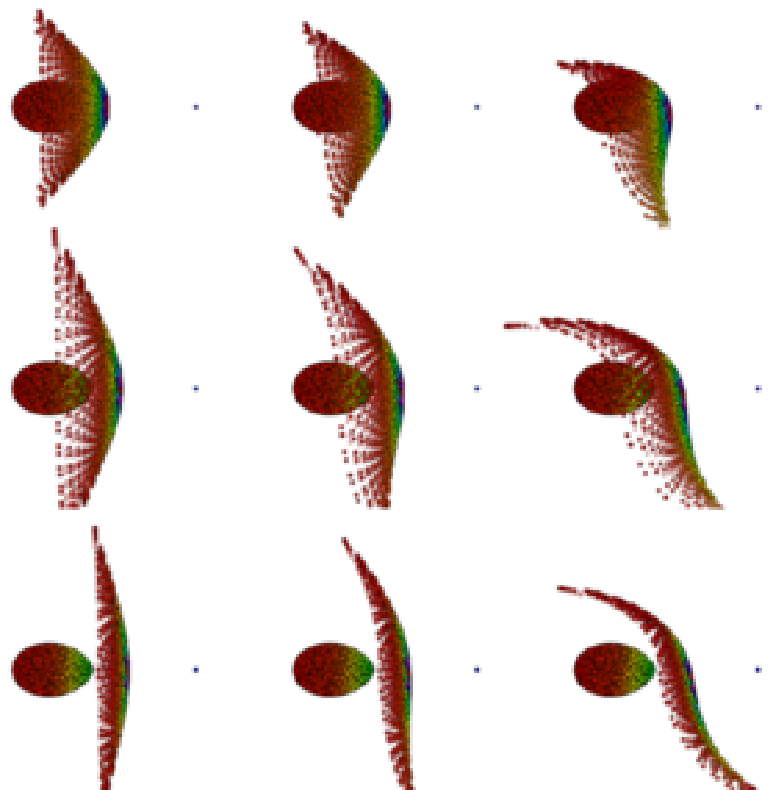}
\includegraphics{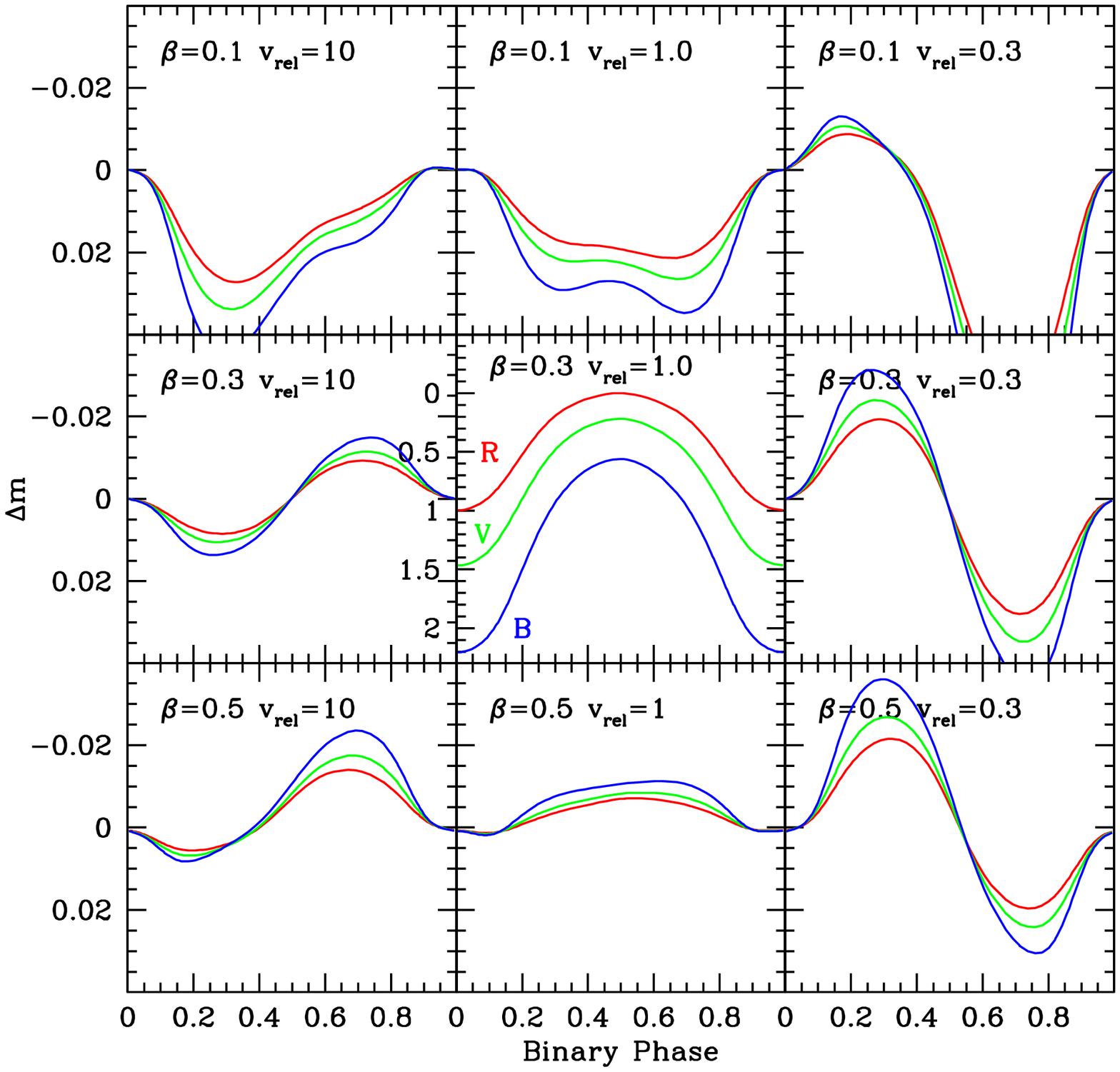}
\begin{center}
\caption{\label{ShsurfLCs} 
Intra-Binary shock geometry dependence on the principal wind parameters. Left:
companion wind momentum increases from top to bottom and velocity decreases 
from left to right. The intrabinary shock is color coded by the fraction of
the shocked power hitting the companion surface and the companion is color coded by the
local heating power. Right: corresponding BVR light curves for $i=70^\circ$. 
The center panel shows the curves for a $\beta=0.3$, $v_{\rm Rel}=1.0$, normalized
to the $R$ maximum. The surrounding panels show light curve changes from this case. 
}
\end{center}
\vskip -0.5truecm
\end{figure*}

\section{Geometry Dependence for the IBS and Companion Light Curves}

	The basic IBS structure is set by $\beta$ and $v_{\rm Rel}$. Small $\beta$
lie close to the companion surface, $\beta=1$ produces a flat, mid-orbit shock, for
large $\beta$ the shock lies close to the pulsar.
Large $v_{\rm Rel}$ produces a nearly symmetric IBS, while for $v_W < v_{orb}$ the sweep-back
is appreciable. In all cases, the most energy is intercepted near the IBS `nose' and
re-radiated to the companion surface. 

Figure 2 shows the basic $\beta$ dependence for a relatively fast companion
wind with $v_{\rm Rel}=3$, with the middle and bottom rows showing the skymap and three light curve
cuts, respectively, for the direct (IBS synchrotron) emission of the
`slow-cooling' tangential component.  For small $\beta$ there
is a substantial effect from eclipse by the companion \citep{boget11}. However, for most
models, the dominant effect is from Doppler beaming at the limb of the pulsar limb shock.
The result is a double-peaked X-ray (synchrotron) light curve centered on optical
minimum (MSP radio eclipse). For large $\beta$ (weak PSR, strong companion wind) Doppler beaming controls
the light curve, which is now centered opposite the MSP radio eclipse. Note that there
is substantial asymmetry even for the the relatively fast companion wind shown here;
for small $v_{\it Rel}$ or small inclination $i$ often only a single peak appears.

In figure 3 we show two example
X-ray light curves. The first is for the original black widow PSR J1959+2048 \citep{huet12}, 
which has a highly energetic pulsar and weak companion wind (small $\beta$). The second,
J2129$-$0429 \citep{robet15} is a long period period, lower ${\dot E}$ redback with a 
relatively massive secondary $\sim 0.4 M_\odot$ undergoing quasi-Roche lobe overflow 
\citep[qRLOF,][]{bet16}. As such it plausibly has a rather high ${\dot M}_W$ and 
thus large $\beta$.
This is similar to the situation seen for high-mass $\gamma$-ray binaries such as LS 5039.

	For $\beta <1$, the prompt post-shock IBS emission illuminates and heats 
the companion. In figure 4 we show the shock geometry (now including varying $v_{\it Rel}$) 
and the optical light curves of the heated companion.  The
resulting heating pattern differences are not visually striking, but they introduce
substantial light curve asymmetry, especially for small $\beta$ and $v_{\rm Rel}$
The examples shown here have $\lambda=0.7$. This parameter has modest effect on the
light curve shape unless $\beta$ is very small. The orbital sweep back for very 
small $v_{\rm Rel}$ can wrap the IBS around the companion; we assume that the portion
of the IBS beyond the tangent point (companion backside) does not intercept pulsar flux
or radiate. As expected, $v_{\it Rel}$ dominates the light curve asymmetry and
the sensitivity to the heating pattern is strongest for blue colors.

\section{Application to PSR J2215+5135}

	PSR J2215+5135 is a redback (RB) system, a $P=2.6$\,ms 
${\dot E}=7.4I_{45} \times 10^{34}{\rm erg\, s^{-1}}$ (with the neutron star moment of
inertia $I_{45}10^{45} {\rm g\,cm^2}$) millisecond pulsar
in a $P_b=4.14$\,hr orbit with a $\sim 0.23 M_\odot$ companion. Schroeder \& Halpern 
(2014, hereafter SH14) obtained high-quality BVR light curves of the companion over many orbits,
finding that the source varies from $V\approx18.7$ to 20.2\,mag, showing strong heating.
Their fit with the ELC code and a photometry table generated from the PHOENIX model atmospheres
\citep{huet13} suggested small inclination $i$ and had a number of peculiarities, including
poor agreement with the observed colors and a highly significant phase shift of optical 
maximum by $\Delta \phi \approx -0.01$ with respect to the radio-pulse ephemeris. \citet{rgfk15}
obtained Keck LRIS spectra throughout the orbit and were thus able to greatly improve the
model fits, finding a much larger inclination $i \approx 90^\circ$ (and hence much smaller
component masses).  \citet{genet14} observed the system in the X-rays with CXO, finding an
X-ray minimum near orbital phase $\phi \approx 0$ (pulsar superior conjunction, optical
minimum, radio eclipse), 
which they interpret as due to variable obscuration of emission from an intrabinary shock 
around the companion. Thus this system is a a good example to test our IBS model and indirect 
re-heating code.

	We use here 103 $B$, 55 $V$, and 113 $R$ magnitudes from SH14. The radio pulsar
timing gives us $x=a_1\, {\rm sin}\,i = 0.468141433$\,lt-s and accurate orbital ephemeris to
phase the photometry points. The basic ICARUS model parameters are the underlying
temperature of the star (actually $T_N$ of the unheated ``night'' face), a heating flux denoted $L_H$, 
the orbital inclination $i$ and the mass ratio $q$. Here
$$
L_H = (T_D^4-T_N^4) 4\pi [x_1(1+q)]^2 \sigma/{\rm sin^2}i
$$
where we have assumed an effective albedo = 0. For a spindown power ${\dot E}$ we may alternatively
write a heating efficiency $\eta=L_H/{\dot E}$. Physically, we expect this to be modest and 
several BW/RB do indeed show $\eta \le 0.1$, however other systems show apparent $\eta \ge 1$
\citep{bet13,rfc15}; we can consider $\eta$ an alternative heating parameter. To match the
observed fluxes the solution also depends on the Roche lobe filling factor $f_1$, the system 
distance, and the interstellar extinction $A_V$. In practice, these last three parameters are
substantially covariant, while having only a weak affect on determinations of $T_c$, 
$i$ and $q$. 

	One challenge to any light curve fitting is the small, but obvious
offset between the timing ephemeris and the time of optical maximum. SH14 quote a heating
center phase shift $\Delta \phi =-0.0140\pm 0.0005$ (ELC fit). In \citet{rgfk15} we found
$\Delta \phi =-0.0089$, with very large statistical significance. Any model that does not
have such a shift is completely unacceptable. This offset, and similar shifts and asymmetries
noted for other BW/RB are prime motivations for an indirect heating model. In our model
the heating asymmetry is introduced via $v_{\rm Rel}$. Note that this is a physical parameter
with a meaningful value, and introduces asymmetry without an arbitrary phase shift. Since
this parameter dominates the INS fit to the optical light curves, it replaces $\Delta \phi$,
leaving the same number of degrees of freedom. More detailed fits (or fits including
X-ray light curves, see below) can include $\beta$ or even $\lambda$, but the optical
dependence is generally weak.

	We thus compare fits with ICARUS-IBS using, as in \citet{rgfk15}, Harris BVR 
color tables from the PHOENIX models tabulated at the Spanish Virtual Observatory 
(svo2.cab.inta-csis.es). We discuss the color sensitivities and then
turn to the fits' dependence on other parameters.
Our results are summarized in Table \ref{Fits} and Figure 5. 

\subsection{$T_D$, Extinction and Color Terms}

	The color as a function of orbital phase should be a powerful constraint
on the heating distribution. Several factors typically complicate its use. First,
there is inevitably some uncertainty in the observations' zero point calibrations.
For example, for the J2215 BVR set, absolute photometry errors may be as large
as 0.1 mag (bootstrap estimate, J. Tan, private communication), although night-to-night
stability suggests that the relative photometry is considerably better. Also, there 
is appreciable degeneracy
between $T_{eff}$ and $A_V$ (and the distance modulus DM) -- in fact for the
$B-V$, $V-R$ colors in our $T_{eff}$ range, the degeneracy is particularly bad. 
If the heating model (or the data calibration) are imperfect, such degeneracy
can allow subtle light curve shape disagreements to pull the fit values to
incorrect temperatures.

	Accordingly it may be useful to use external (non-photometric) constraints
to control some model parameters. For example RGFK15 find effective
temperatures $T_D \approx 9000$\,K and $T_N\approx 6000$\,K from the spectroscopy.
This implies a nominal heating efficiency $\eta_H=L_H/{\dot E} \approx 0.48/I_{45}$ 
for a direct radiative heating
picture.  Constraints on the extinction are less direct.  The pulsar dispersion measure
DM=69.2${\rm cm^{-3}}$pc, which corresponds to a distance $d=3.0 \pm 0.35$\,kpc in the NE2001
DM model, converts to ${\rm N_H=2.07^{+0.9}_{-0.6} x 10^{21}cm^{-2}}$ \citep{hnk13}
which corresponds to ${\rm A_V=0.72}$ \citep{fgos15}. However, more direct estimates from 
Pan-STARRS photometry \citep{gsfet15} give ${\rm A_V=0.47}$ at 3\,kpc and a 
maximum Galactic ${\rm A_V=0.71}$ in this direction. Unfortunately the existing X-ray 
exposure is too short to give a constraining absorption measurement. Thus we conclude
that ${\rm A_V\le 0.7}$.

	In practice we find that unconstrained fits to the SH14 photometry give 
$T_D \approx$13,000\,K and $A_V\approx 1.4$ without IBS heating. This indicates
that the measured colors are not consistent with the standard atmosphere models,
or that the light curve shapes drive the model to artificially high $T_D$. We did attempt
to see what zero point shifts could drive the best-fit $T_D$ down to 9000\,K, but the
large $\sim 0.1$\,mag required values resulted in rather poor fits to the light curve shapes.

\begin{deluxetable*}{llrrrrr}[t!!]
\tablecaption{\label{Fits} ICARUS Model Fits}
\tablehead{
\colhead{Param.} & \colhead{Direct\tablenotemark{a}}& \colhead{Direct-$T_D$\tablenotemark{a}} & \colhead{IBS} & \colhead{IBS-$T_D$} 
}
\startdata
$i$       &79.6$\pm$3.2   &78.3$\pm$3.1   &89$\pm$8       &83$\pm$6   \cr
$f_1$     &0.867$\pm$0.004&0.912$\pm$0.004&0.852$\pm$0.006&0.905$\pm$0.004  \cr
$T_N$(K)  &7670$\pm76$   &6421$\pm23$     &7290$\pm79$    &6416$\pm58$      \cr
$L_H(10^{34}{\rm erg/s})$      
          &29.0$\pm 3.5$ &5.7$\pm 0.2$   &32.9$\pm$3.0    &11.4$\pm$0.8   \cr
$T_D$(K)  &13,350\tablenotemark{b} &9000\tablenotemark{c} &11,710\tablenotemark{b} &9000\tablenotemark{c} & \cr
$A_V$     &1.39$\pm0.03$ &0.75$\pm0.02$  &1.25$\pm$0.03   &0.79$\pm0.04$  \cr
DM (mag)  &13.76$\pm$0.03&13.56$\pm$0.01 &13.62$\pm$0.02  &13.51$\pm$0.03 \cr
dB/dR     &0.03/0.00&0.06/0.01           &0.04/0.00       &0.06/0.00\cr
$v_{\rm Rel}$ &--            &--             &0.277$\pm$0.007 &0.283$\pm$0.007\cr
$\chi^2$  &876           &1228           &729             &1018  \cr   
\enddata

\tablenotetext{a}{ Best (arbitrary) phase shift $\delta \phi_B= -0.01$ applied.}
\tablenotetext{b}{ Flux weighted Effective $T_D$ at $\phi_B=0$ computed from $L_H$ etc.}
\tablenotetext{c}{ Flux weighted Effective $T_D$ at $\phi_B=0$ fixed.}
\end{deluxetable*}

\subsection{Fitting Results}

	In all cases, we find that the fits improve if we add a small positive offset to 
the measured B magnitudes; in one case a 0.01mag relative shift in R is also indicated.
Fitting with a symmetric direct heating model is completely unacceptable,
with a minimum $\chi^2 =2400$. Thus at a minimum, we must introduce the (here arbitrary)
phase shift of the heating center by $\Delta \phi=-0.01$. This gives a much more acceptable
$\chi^2 =876$ (Table 1). However, including the IBS reprocessing of the heating 
(with $\beta=0.5$) further improves the fit to $\chi^2 =729$; the model parameters are 
listed in Table 1 and the model
and fit residuals are shown in Figure 4. This best fit is however, not statistically
acceptable since with 271-7=264 degrees of freedom, this is $\chi^2/DoF = 2.76$. 
Certainly a substantial portion of this large $\chi^2$ is caused by individual outlier
points. Also, the scatter in Figure 5 is larger than expected from the error flags, 
especially near
minimum, which indicates that either the photometric errors of SH14 are underestimated
or that there is true stochastic photometric variability. However, inspection of the residuals,
especially for V and B, also shows systematic trends. Clearly our IBS heating model
does not give a perfect representation of the true surface temperature distribution.
For example the positive V and negative B residuals around maximum suggest that the
nose heating is overestimated by the IBS pattern. Accordingly, the quoted fit statistical
errors, even inflated by $\chi^2/DoF$, are not a complete description of the uncertainties,
and some amendment to the model is needed.

	As noted in \S 4.1, the incompleteness of the model is also seen from
the preference in both the best direct and best IBS models for large $T_{\rm eff} \approx 12,000
-13,500$\,K and $A_V \approx 1.3$. One worries that the unmodeled effects pulling the
$T_D$ to such large values may also bias other parameters. Accordingly we also fit
while constraining $T_D=9000$\,K.  The $\chi^2$ values increase, of course, but the 
best fit baseline temperature $T_N$, heating luminosity and $A_V$ drop to more reasonable
values. The primary geometrical difference is somewhat larger Roche lobe fill factor $f_1$.

	Note that the IBS fits have significantly lower $\chi^2$ than the direct
heating models, both with and without $T_D$ constraint. These are large
$\Delta \chi^2 > 150$ changes relative to the $\Delta \chi^2 \approx 6-10$ associated
with the parameter error ranges. These $1\sigma$ ranges are projected, multi-parameter
errors in all cases.  In these fits we have
held fixed $\beta$=0.5 and $\lambda$=0.7, since the optical light curves depend only
weakly on these quantities, and have varied only the $v_{\rm Rel}$ IBS parameter. Since
this parameter replaces the arbitrary phase shift $\Delta \phi$ required by the direct 
heating models, the fits all have effectively the same number of degrees of freedom.
If we do free $\beta$ in the IBS fits, we find $\beta=0.75 \pm 0.3$ and 
 $\beta=0.49 \pm 0.5$ for the IBS and IBS$-T_D$ cases, respectively. As expected
for such poorly determined parameters, the $\chi^2$ decrease is small.

\begin{figure*}[t!!]
\vskip 7.4truecm
\includegraphics{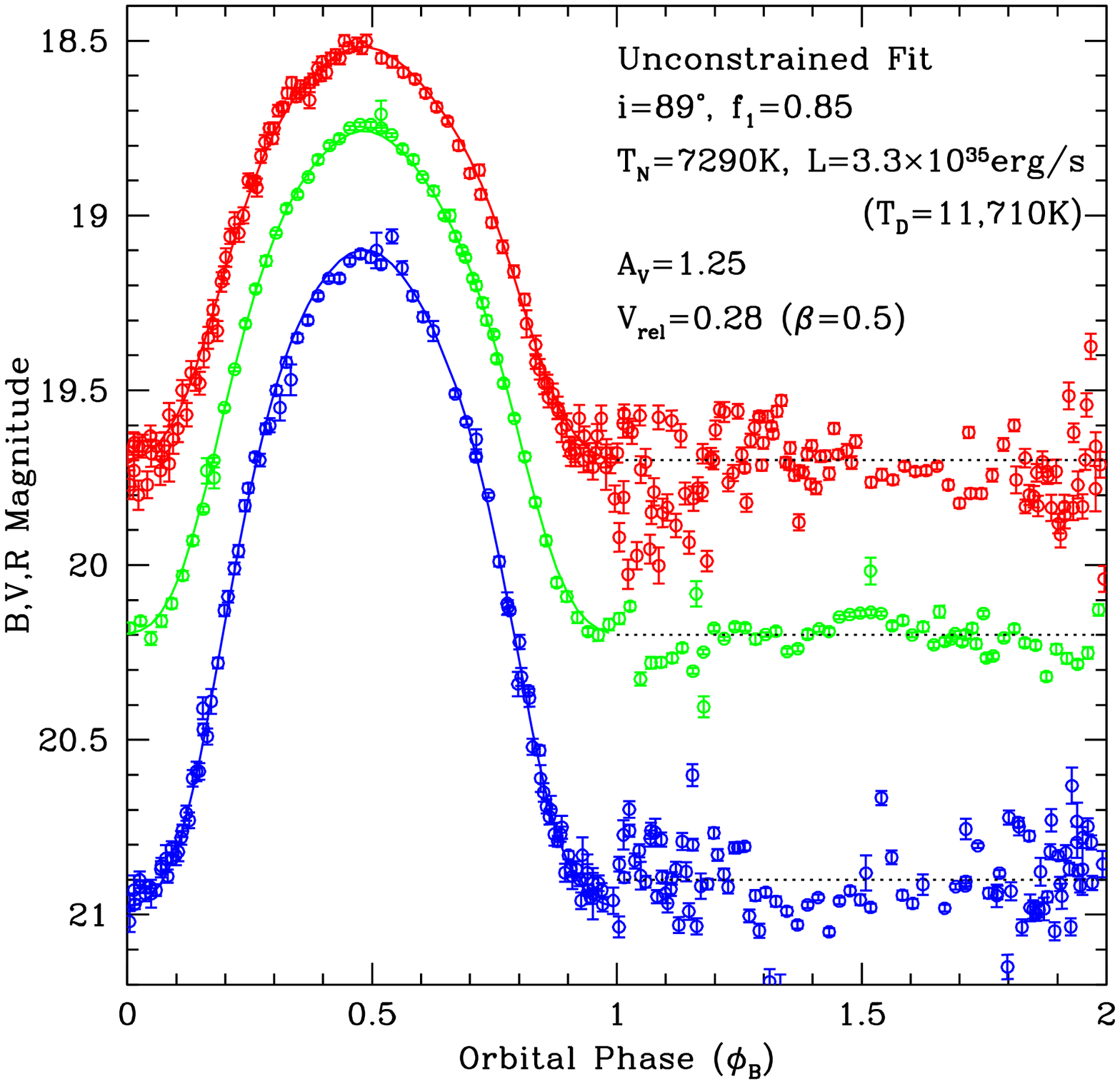}
\includegraphics{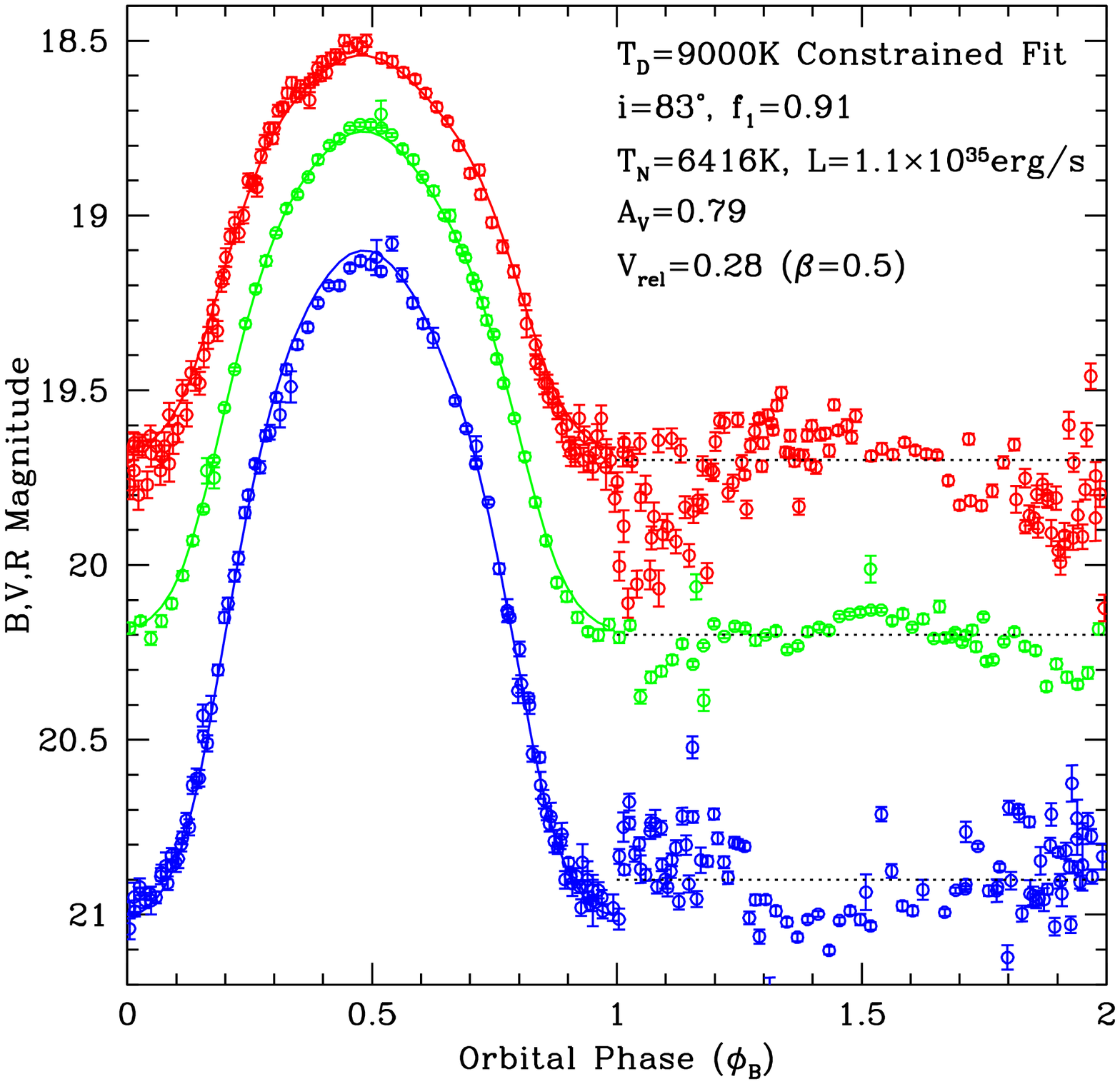}
\begin{center}
\caption{\label{Fitplot} 
SH14 BVR light curves with IBS-fit model (1st period). The second period shows 
the fit residuals (expand $3\times$). Left: best fit model, Right: best fit $T_D$ constrained model.
}
\end{center}
\vskip -0.8truecm
\end{figure*}

	Realistically, the best constraints on the IBS parameters will, in many cases, come from
X-ray orbital light curves. In the case of J2215, we have only a low statistics 17\,ks ACIS
exposure to compare with. Figure 6 shows the X-ray count rate, phased with pulsar superior
conjunction at $\phi=0$ along with a curve computed for the IBS tangential emission for the
parameters fit to the optical light curves. Although these X-ray data are not used in
the fit it is encouraging
that for these parameters the model predicts a single strong X-ray peak at $\phi \approx 0.25$,
in excellent agreement with the data. A high quality X-ray light curve would be very
useful to directly constrain the IBS geometrical parameters, as well as measure 
the radiation spectrum.

\begin{figure}[h!!]
\vskip 8.8truecm
\includegraphics{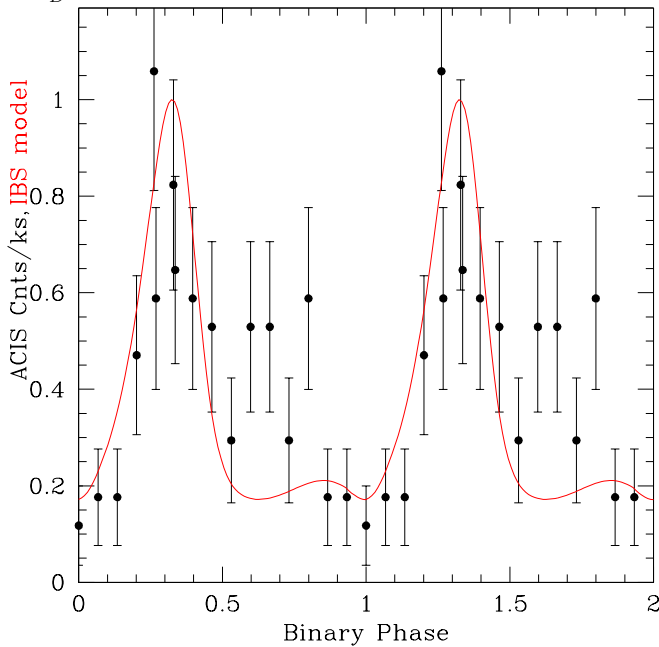}
\begin{center}
\caption{\label{Fitplot} 
J2215+5135 ACIS X-ray light curve (points) with the predicted IBS tangential synchrotron
light curve for the parameters of the model which best fit the optical data.
}
\end{center}
\vskip -0.8truecm
\end{figure}

\section{Conclusions}

	We have investigated a model in which pulsar radiation is reprocessed through
an intrabinary shock before heating a low mass companion. Our shock geometry is controlled
by two dimensionless parameters, $\beta$ and $v_{\rm Rel}$, and is idealized as following
the contact discontinuity in the thin-shock limit. When we compute the expected synchrotron
emission from the shocked pulsar wind accelerated tangential to the contact discontinuity, 
we find a variety
of asymmetric X-ray light curves, dependent on these two parameters. The companion heating,
modeled as a result of direct companion illumination from the prompt post-shock emission
also produces a range of asymmetric light curves.

	In applying these models to PSR J2215+5135, which has well measured optical
light curves and spectra, we find that the IBS model fits the data appreciably better
than the direct illumination model, even when the latter is allowed an arbitrary phase shift.
This improvement persists even we constrain the fits to match the spectrally
determined $T_D$ and the expected $A_V$ in this direction. Thus the model presents a
substantially improved representation of the data -- and provides a physically
meaningful origin
of the observed phase shift of the optical maximum. In addition, the observed X-ray light curve,
while of too low statistics to allow a detailed parameter fit, does provide a good
match to the IBS light curve expected from the best-fit to the optical data. This is all
very encouraging and suggests that $v_{\rm Rel}$ is a relatively low 0.28. $\beta$ is larger
at $\approx 0.5$ but is not well determined. These nominal values imply a velocity
(relative to the companion center, at Roche lobe exit) of $115$\,km/s and a companion
mass loss rate of 
$$
{\dot M} \approx \beta {\dot E}/(c v_{orb} v_{\rm Rel}) =
$$
$$ 
\qquad (\beta {\dot E} P_B {\rm sin} i )/(2\pi c^2  x v_{Rel})
= 1.7 \times 10^{-9} I_{45} M_\odot{\rm y^{-1}}
$$
Thus, the shape and sweepback of this
IntraBinary Shock implies that, at the present companion mass-loss rate, evaporation 
has a characteristic timescale $\tau_{\rm evap} = M_c/{\dot M} \approx 150$Myr. If this rate persists,
we expect J2215 to be an isolated MSP in $<$Gy. It should be noted, however, that
J2215 has parameters rather similar to those of the `transitioning' MSP J1023+0038
and so might spend time in an accretion phase, suppressing pulsar irradiation
and leading to a lower mass loss rate. The relatively large $\beta$ preferred
by our model fits, suggest that the pulsar wind could indeed be overwhelmed by
a fluctuation in mass-loss rate, burying the pulsar.

	Despite these successes, the model is clearly not complete, as shown by the
large residual $\chi^2$. Some of this is due to understated errors,
and individual outlier points (possibly indicating flare events as seen, e.g. for
PSR J1311$-$3430, Romani et al 2105). However, systematic light curve shape residuals,
especially for models constrained to match the spectral temperature, indicate 
deficiencies in the computed heating pattern. Also, although zero-point errors in
the photometry undoubtedly play a role, the preference of the fits for high $T_D$
indicates an incorrect heating distribution. Finally, the IBS, while capturing
a much larger fraction of the pulsar spin-down power, does not focus this power to
the companion surface. Indeed, our assumed prompt radiation, computed with the forward
shock jump conditions, takes the incident pulsar power and deflects it away
from the shock normal (and thus, for most positions on the IBS, further from the companion).
Thus we find that re-processing the pulsar power through the IBS requires up to
$2\times$ {\it larger} pulsar luminosity (for this prompt post-shock illumination
picture) than direct heating. For the best-fit IBS model we infer an efficiency
$\eta = 4.4/I_{45}$ and for the $T_D$-constrained IBS model $\eta = 1.5/I_{45}$. These
factors are for a ${\rm sin^2}$ wind flow and are only modestly reduced for a ${\rm sin^4}$
distribution. Finally, while we have matched the apparent asymmetry for J2215, some
other wind-driving pulsars display much larger optical heating asymmetries, that 
would be difficult to produce in this model even with small $v_{\rm Rel}$.

	We conclude that while IBS-reprocessing through a swept back model can be
a viable solution for some wind-driving pulsars, an additional physical ingredient
is likely needed to fully match the heating data and to explain particularly extreme cases.

The most likely culprit is ducting by companion magnetic fields. There is in fact good
reason to believe that substantial fields can be supported by the companion. 
BW and RB are short period, tidally locked binaries so the secondaries are, by definition, 
rapidly rotating stars. Also, since night side temperatures of black widows and redbacks
are appreciably higher than expected for the unperturbed star, rapid motion must be advecting
heat to the night sides. These convective motions in the presence of rapid spin give a
plausible dynamo origin for large, dynamic B fields. 

	With a typical standoff distance $r_0\sim 0.3R_\odot \sim 2\times 10^{10}{\rm cm}<a$,
companion-supported fields with a dipole of strength $B_C$ and coherence scale
$r_C$ can channel the wind flow if
$$
[B_C(r_0/r_C)^3]^2/8\pi > {\dot E}/(4\pi a^2 c)
$$
or $B_C > 8 (r_0/r_C)^3$G. So the dominance depends on the large scale coherence
of the companion dipole field located near $r_C(0)$. For J1311$-$3430
\citet{rfc15} observed apparently magnetically-driven flares with the surface heating and
flux giving a characteristic size $\sim r_C(0)/3 \sim 0.02 a \sim 0.1 r_0$
and energy density equivalent to $B_C \sim 10$kG, which could be dynamically
significant at the IBS standoff distance. If similar field exist in the companions
of other BW/RB we may expect them to redirect the energy released in an IBS. We can
then imagine IBS particles precipitating from a cross sectional area $\sim \pi r_0^2$
to the companion surface, inducing heating at the field line foot points. 
Such local heating would inevitably induce temperatures higher than the mean
$T_D$ and will, in general, have foot point hot-spots offset from the sub-pulsar point.
Indeed if the local field is not largely dipolar we may have precipitation 
at many poles and a complex, offset heating pattern. It remains to be seen if
a detailed model of such field-mediated heating, which could be fit to observed
light curves and spectra, could have useful predictive power.

\bigskip
\bigskip

We thank Hongjun An and Yajie Yuan for helpful discussions about shock interactions,
Josh Tan for insight into the SH14 data  and Rene Breton for advice on the ICARUS code.


\begin{thebibliography}{}

\bibitem[Aldcroft, Romani \& Cordes(1992)]{arc92}Aldcroft, T., Romani, R.~V. \& Cordes, M. 1992, ApJ, 400, 638
\bibitem[Bellem et al.(2016)]{bet16}Bellm, E.~C., Kaplan, D.~L., Breton, R.~P. et al. 2016, ApJ, 816, 74
\bibitem[Bosch-Ramon, Barkov \& Perucho(2015)]{bbp15}Bosch-Ramon, V. Barkov., M.~V. \& Perucho M. 2015, AA, 577, 89 
\bibitem[Breton et al.(2013)]{bet13}Breton, R.~P., et al. 2013, ApJ, 769, 108
\bibitem[Bogdanov et al.(2011)]{boget11}Bogdanov, S., Archibald, A.~M., Hessels, J.~W.~T., et al. 2011, ApJ, 742, 97
\bibitem[Bogovalov et al.(2012)]{boget12}Bogovalov, S.~V., Khangulya, K., Koldoba, A.~V., Ustyugova, G.~V. \& Aharonian, F.~A. 2012, MNRAS, 419, 4326
\bibitem[Callanan, van Paradijs \& Rengelink(1995)]{cvr95}Callanan, P.~J., van Paradijs, J. \& Rengelink, R. 1995, ApJ 439, 928
\bibitem[Cant\'{o}, Raga \& Wilkin(1996)]{crw96}Canto', J., Raga, A.~C., \& Wilkin, F.~P. 1996, ApJ, 469, 729.
\bibitem[Djorgovski \& Evans(1988)]{de88}Djorgokski, S., \& Evans, C.R. 1988, ApJ 335, L61
\bibitem[Foight et al(2015)]{fgos15}Foight, D.~R., Guever, T., Oezel, F. \& Slane, P.~O. 2015, ArXiv150407274
\bibitem[Gentile et al.(2014)]{genet14}Gentile, P.~A., Roberts, M.~S.~E., McLaughlin, M.~A., et al. 2014, ApJ, 783, 69
\bibitem[Green et al.(2015)]{gsfet15}Green, G.~M., Schlafly, E.~F., Finkbeiner, D.~P. et al. 2015, ApJ, 810, 25
\bibitem[He, Ng \& Kaspi(2013)]{hnk13}He, C., Ng, C.-Y., \& Kaspi, V.~M. 2013, ApJ, 768, 64.
\bibitem[Huang et al.(2012)]{huet12}Huang, R.~H.~H., Kong, A.~H.~K., Takata, J., et al. 2012, ApJ, 760, 92
\bibitem[Husser et al.(2013)]{huet13}Husser, T.-O., Wende-von Berg, S., Dreizler, S., et al. 2013, AA, 533, A6
\bibitem[Komissarov \& Lyutkov(2011)]{kl11}Komissarov, S~S., \& Lyutikov, M. 2011, MNRAS, 414, 2017
\bibitem[Orosz \& Hauschildt(2000)]{oh00}Orosz, J. A., \& Hauschildt, P. H. 2000, AA, 364, 265
\bibitem[Parkin \& Pittard(2008)]{pp08}Parkin, E.~R., \& Pittard, J.~M. 2008, MNRAS, 388, 1047
\bibitem[Roberts et al.(2014)]{robet14}Roberts, M.~S.~E., McLaughlin, M.~A., Gentile, P.~A. et al. 2014, AN, 335, 315
\bibitem[Roberts et al.(2015)]{robet15}Roberts, M.~S.~E., McLaughlin, M.~A., Gentile, P.~A. et al. 2015, ArXiV, 1502.07208
\bibitem[Romani(2015)]{r15}Romani, R. W. 2015, ApJ, 812, L24 
\bibitem[Romani, Filippenko \& Cenko(2015)]{rfc15}Romani, R.~W., Filippenko, A. V., \& Cenko, S. B. 2015, ApJ, 804, 115.
\bibitem[Romani et al.(2015)]{rgfk15}Romani, R.~W., Graham, M.~L., Filippenko, A. V., \& Kerr, M. 2015, ApJ, 809, 10
\bibitem[Schroeder \& Halpern(2014)]{sh14}Schroeder, J., \& Halpern, J. P. 2014, ApJ, 793, 78
\bibitem[Stappers et al.(2001)]{stap01}Stappers, B.W, van Kerkwijk, M. H., Bell, J.F. \& Kulkarni, S. R. 2001, ApJ, 548, 183
\bibitem[Tchekhovskoy, Spitkovsky \& Li(2013)]{tsl13}Tchekhovskoy, A., Spitkovsky, A. \& Li, J.~G. 2013, MNRAS, 435, L1
\bibitem[van Kerkwijk, Breton, \& Kulkarni(2011)]{vKBK11}van Kerkwijk, M. H., Breton, R. P., \& Kulkarni, S. R. 
2011, ApJ, 728, 


%


\end{thebibliography}
\end{document}